\documentstyle[aaspp4]{article}
 
\lefthead{Elmegreen}
\righthead{Star Formation in One Crossing Time}
\slugcomment{{\it Astrophysical Journal}, 530, Feb 10, 2000, in press}
\begin{document}
 
\title{Star Formation in a Crossing Time}
 
\author{Bruce G.~Elmegreen\altaffilmark{1}}
\altaffiltext{1}{IBM Research Division, T.J. Watson Research Center,
P.O. Box 218, Yorktown Heights, NY 10598}
 
\begin{abstract} Observations suggest that star formation occurs in only
one or two crossing times for a range of scales spanning a factor of
$\sim1000$. These observations include (1) measurements of embedded
cluster ages in comparison with the cloud core dynamical times, (2)
measurements of the age difference versus separation for clusters in the
Large Magellanic Clouds in comparison with the crossing time versus
size correlation for molecular clouds, (3) hierarchical structure of
embedded young clusters, and (4)
the high fraction of dense clouds that contain star formation.

Such a short overall time scale for star formation implies that sources
of turbulent energy or internal feedback are not required to explain or
extend cloud lifetimes, and that star and protostar interactions cannot
be important for the stellar initial mass function. Stars appear
in a cloud as if they freeze out of the gas, preserving the
turbulent-driven gas structure in their birth locations. The galaxy-wide
star formation rate avoids the Zuckerman-Evans catastrophe, which has
long been a concern for molecular clouds that evolve this quickly,
because the multifractal structure of interstellar gas ensures that only
a small fraction of the mass is able to form stars. Star formation on
large scales operates slower than on small scales, but in most cases the
whole process is over in only a few dynamical times.

\end{abstract}

\section{Introduction}

The duration of star formation in a region of a particular size is a
physical parameter related to the star formation process that has rarely
been considered interesting, except perhaps for the notion that star
formation lasts sufficiently long in GMCs that sources of
turbulent cloud support and internal feedback are necessary. This
concept of inefficient and prolonged star formation goes back to the
first galactic surveys when it was realized (Zuckerman \& Evans 1974;
Zuckerman \& Palmer 1974) that the Galactic CO mass and density are too
large, and the total star formation rate too small, to have the
conversion of gas into stars take place on anything shorter than several
tens of crossing times per cloud. Early claims to the longevity of
clouds were also based on chemical abundances and slow reaction rates at
the average cloud density. Even the first measurement of an age for
GMCs, of around $3\times10^7$ years from cluster disruption of gas
(Bash, Green, \& Peters 1977) was $\sim40$ dynamical times at what was
believed to be the average GMC density of $\sim10^3$ cm$^{-3}$. This
time was comparable to the shortest theoretical lifetime estimated from
internal clump collisions (Blitz \& Shu 1980), but it was still long
compared to the dynamical time. 

As a result of these ideas, astronomers have been trying for two decades
to understand how self-gravitating clouds can support themselves for
many dynamical times, considering the common belief (Goldreich \& Kwan
1974; Field 1978) that supersonic turbulence should dissipate more
rapidly. Two types of models arose, those in which turbulence supports a
cloud on the large scale but not on the small scale (Bonazzola et al.
1987; Leorat, Passot, \& Pouquet 1990; Vazquez-Semadeni \& Gazol 1995),
and those in which stellar winds continuously drive turbulence to
support a cloud (Norman \& Silk 1980), possibly with some type of
feedback to maintain stability (Franco \& Cox 1983; McKee 1989).

A variety of recent observations now suggests a different picture.
Star formation appears to go from start to finish in only one or two 
crossing times on every scale in which it occurs (Sect. \ref{sect:obs}).
The star formation rate does not just {\it scale} with the self-gravity
rate, as recognized for a long time, it essentially
{\it equals} the self-gravity
rate. The first hint at such quick star formation came long ago from the
observation that a high fraction of clouds contain stars (Beichman et
al. 1986), combined with the idea that pre-main sequence lifetimes are
short. As a result, the total cycle of star formation, before and after
cloud dispersal, has to be short as well (Elmegreen 1991). 

Today the observational picture for rapid star formation is more
complete, consisting of direct measurements, such as embedded cluster
ages or cluster age differences in comparison to the associated gas
turbulent crossing times, and indirect indicators such as hierarchical
structure in embedded stellar groups and the fraction of clouds with
embedded stars. Much of this data is well known, but it has not been
viewed together in this fashion, and it is rarely interpreted as an
indication that star formation occurs in only one crossing time. The
implications of such an interpretation could be important for
understanding the processes of star formation.

Here we point out that star formation in a crossing time implies four
changes to way we view the physical processes involved: (1)
Feedback and cloud support from turbulence are not necessary for
molecular clouds. (2) Protostar interactions have too little time to
affect the average stellar initial mass function, which must be
determined primarily from a rapid sampling of existing cloud structure rather
than from a long time sequence of internal cloud dynamics.  (3) The
chemical clock inside a molecular cloud is determined by transient,
high-density events, rather than by slow chemistry at the mean density.
(4) The inefficiency of star formation on a Galactic scale results from an
{\it inability} of most molecular or CO-emitting gas 
to form stars at all, and not from any of the previous
explanations, which include a delay in the onset of star
formation, slow magnetic diffusion, turbulent cloud support, local
inefficiencies in a cloud core, and cloud disruption. 
This inability to form stars arises simply from the turbulence-driven
structure of clouds (Falgarone, Phillips \& Walker 1991): 
most molecular gas is either at too low a density, in an
interclump medium, or it is dense, evanescent, and too small to be
self-gravitating (Padoan 1995).

\section{Observations of rapid star formation}
\label{sect:obs}

\subsection{Age difference versus size for LMC clusters}
\label{sect:lmc}

A correlation between the age difference and spatial separation for
Cepheid variables in the LMC suggests that star formation lasts for a
time that is approximately equal to the dynamical crossing time for a
wide range of scales (Elmegreen \& Efremov 1996). This implies not
only that young star positions are hierarchical, as had been claimed
before (Feitzinger \& Braunsfurth 1984; Feitzinger \& Galinski 1987),
but also that the timing for star formation is hierarchical, with many
small regions coming and going in the time it takes the larger region
surrounding them to finish. A second study using cluster ages and
positions in the LMC confirmed this Cepheid result (Efremov \& Elmegreen
1998), and illustrated again that the time scale for coherent star
formation in a region is always about one turbulent crossing time,
scaling approximately with the square root of size.

Figure 1 shows this result by plotting the crossing times inside
molecular clouds and sub-clouds of various sizes $S$ versus these sizes
(which are essentially cloud diameters). The crossing time is defined as
the half-size (radius) divided by the Gaussian dispersion in internal
velocity, $c$. The data is form the literature, as indicated. Superposed
on this is the analogous correlation between age-separation and
distance-separation for $244$ LMC clusters in the age range from 10 to
100 million years (from Efremov \& Elmegreen 1998). The clusters lie on
a continuation of the crossing time-size relation for individual clouds,
suggesting that in each region in which this cluster hierarchy is
observed, the {\it duration} of the star formation process is an average
of about one crossing time. 

Hierarchical clustering in time and space is also shown by cluster
pairs, analogous to h and $\chi$ Persei in our Galaxy. Equal-age
pairs have been studied in the LMC by Bhatia \& Hatzidimitriou (1988),
Kontizas et al. (1989), Dieball \& Grebel (1998), and Vallenari et al.
(1998) and in the SMC by Hatzidimitriou \& Bhatia (1990). Their
existence implies that star formation is synchronized in neighboring
regions, which means there is only a short time interval available for
the complete formation of a cluster and its neighbor.

\subsection{Substructure in Embedded Infrared Clusters}

A second indication that star formation is extremely rapid is the
observation that some embedded IR clusters have sub-clustering. For
example, IC 342 contains 8 smaller subclusters inside of it with 10 to
20 stars each Lada \& Lada (1995). Such sub-clustering would be mixed up
by star-star scattering and gravitational tidal interactions if the
individual stars had enough time to orbit even once through the cloud
core. Instead, the cluster seems to have crystalized instantly,
preserving the pre-stellar hierarchical cloud structure in the pattern
of young stars. The star formation process is not just beginning in this
region either. At the present time, a fraction equal to about 50\% of
the total cloud mass has already been converted into stars (Lada \& Lada
1995). This fraction is comparable to the likely final efficiency for
the cluster, so the total star formation process is nearly over. 

Other clusters with hierarchical subclustering include NGC 3603
(Eisenhauer et al. 1998), W33 (Beck et al. 1998), and NGC 2264 (Piche
1993), which has two levels of hierarchical substructure, i.e., two main
clusters with two subclusters in one and three in the other. Elson
(1991) found spatial substructure in 18 LMC clusters, and suggested it
might result from merging subclusters. Strobel (1992) found age
substructure in 14 young clusters, and Persi et al. (1997) found both
age and positional substructure in G 35.20-1.74.

Some of the structure inside a cluster could be the result of triggering
(Elmegreen \& Lada 1977), but this operates on a crossing time for the
outer scale too. For example, the subgroups in OB associations listed by
Blaauw (1964), some of which may be triggered by older subgroups, have
spatial separations on the order of $\sim10$ pc, and age differences on
the order of $\sim3$ million years. These numbers fit on the
correlation in figure 1.

\subsection{Statistical Considerations}

If a high fraction of clouds contains stars and the stellar ages are
always young, then the whole star formation process must be rapid. Three
new compilations of this statistical measurement point to this
conclusion. Fukui et al. (1999) finds that about 1/3 of the clusters in
the LMC younger than 10 Myr are associated with CO clouds, while
essentially none of the clusters older than this are significantly
associated. This implies the entire cluster formation process, including
cloud formation and dispersal, is only around 3 Myr. If the average CO
cloud density at the threshold of their detection is in the usual range
from $10^2$ to $10^3$ cm$^{-3}$, then 3 Myr is 2 to 1.5 dynamical times,
respectively (we take a dynamical time to be
$\left(G\rho\right)^{-1/2}$).

Jessup \& Ward-Thompson (1999) found that the mean pre-stellar lifetime
decreases with increasing density, from about $10^7$ yr at $10^3$
cm$^{-3}$ to $5\times10^5$ yr at $3\times10^4$ cm$^{-3}$. At $10^3$
cm$^{-3}$, this time is 5 dynamical times, and at $3\times10^4$
cm$^{-3}$, it is 1.4 dynamical times. 

Myers (1999) confirms the result of Jessup \& Ward-Thompson (1999) using
different data, and finds that the mean waiting time for star formation
begins to decrease rapidly with increasing density once the density
reaches $\sim10^4$ cm$^{-3}$. At that density, the mean waiting time is
1 Myr or less, which is less than $2$ dynamical times. 

\section{Direct pre-main-sequence age measurements}

The age spread for 80\% of the stars in the Orion Trapezium cluster is
apparently less than 1 My (Prosser et al. 1994). The same is true for
L1641 (Hodapp \& Deane 1993). The age spread is much shorter for a large
number (but not necessarily a large fraction) of stars in NGC 1333
because of the preponderance of jets and Herbig-Haro objects (Bally et
al. 1996). In NGC 6531 as well, the age spread is immeasurably small
(Forbes 1996). These short time scales are all less than a few crossing
times in the cloud cores. 

In a recent study of the time history of star
formation in the trapezium cluster, Palla et al. (1999) found 
that most of the low mass stars formed in the last $\sim1$ My, and that 
the rate increased to this value somewhat
gradually before this, perhaps as the associated cloud
contracted.  A comparison of their figures 1 and 3, along with 
their figure 6, indicates that the low mass stars
mostly formed between $10^5$ and $10^6$ years ago. 
The stellar density in the trapezium is now about
$10^3$ M$_\odot$ pc$^{-3}$ (Prosser et al. 1994;
McCaughrean \& Stauffer 1994), so if the local efficiency of
star formation was around 50\% to make a
nearly-bound cluster, then the
prior gas density in the core 
was $\sim6\times10^4$ H$_2$ cm$^{-3}$. This is a
reasonable value considering the densities in other Orion
cluster-forming regions (Lada 1992). The corresponding
dynamical time scale is $\left(G\rho\right)^{-1/2}\sim 0.3$ My.
which is comparable to the isochrone times of the low mass stars. 
The increase in the rate of star formation
during cloud contraction is what should be expected if this rate
always follows the local dynamical rate (Palla et al. 1999), 
because that increases
too during cloud contraction. 

On larger scales, the age spread in a whole OB association is about 10
My (Blaauw 1964), and the prior gas 
mass ($\sim2\times10^5$ M$_\odot$) inside a
typical radius ($\sim20$ pc) corresponds to an average density of
$\sim200$ atoms cm$^{-3}$; this gives a similar 
dynamical time of 6.3 My. On
even larger scales, the age spread in a star complex like Gould's Belt
is $\sim40 $ My (P\"oppel 1997). These larger regions form inside and downstream
from spiral arms in $\sim500$ pc-size cloud complexes that contain
$10^7$ M$_\odot$ (Elmegreen \& Elmegreen 1987; Efremov 1995). The
average density is $\sim5$ atoms cm$^{-3}$, so the dynamical time is
$\sim40$ My.  Note that the large-scale star-forming regions
contain smaller scale regions inside them, and that all of the
regions form on a local dynamical time. This means that several smaller
regions come and go throughout the larger region 
during the time the larger region exists (Elmegreen \& Efremov 1996).

Evidently, the total duration of star
formation in most clouds is only 1 and 2 dynamical times once star
formation begins, and this is true 
for scales ranging from 1 pc to $10^3$ pc. The general 
concept that the star formation time should scale
with the dynamical time is not new, but direct observations of the
actual timescales have been available only recently. 

Some clusters have larger age spreads than the dynamical time, but
this could be the result of multiple bursts. Hillenbrand et al.
(1993) found that the most massive stars (80 M$_\odot$) in NGC 6611 have
a 1 My age spread around a mean age of $\sim2$ My, which is consistent
with the spreads mentioned above, but there are also pre-main sequence
stars in the same region, probably much younger, and a star of 30
M$_\odot$ with an age of 6 My. The LMC cluster NGC 1850 has an age
spread of 2 to 10 My (Caloi \& Cassatella 1998), and NGC 2004 has both
evolved low mass stars and less evolved high mass stars (Caloi \&
Cassatella 1995). In NGC 4755, the age spread is 6 to 7 My, based on the
simultaneous presence of both high and low mass star formation (Sagar \&
Cannon 1995). 

The large age spreads may result from multiple and independent star
formation events, perhaps in neighboring cloud cores or triggered
regions. A merger event or projection effects could disguise the initial
multiplicity. If this is the case, then the relevant dynamical time for
comparison with the age spread should be calculated with the average
density of the whole region surrounding the two cores and not the density
of each. Thus, the whole region could form in less than a few crossing
times, but the currently dense part of the cluster would have too short a
crossing time for the mixture. This consideration of the average density
surrounding multiple clusters is also necessary to explain the large-scale
correlation between duration and size for star-forming regions defined
by Cepheids and clusters in the LMC (cf. Sect. \ref{sect:lmc}).

A good example of this multiplicity may be the Pleiades cluster,
which has the largest reported age spread of any of the well-studied
clusters. Features in the luminosity function (Belikov et al. 1998) and
synthetic HR diagrams (Siess et al. 1997) suggest continuous star
formation over $\sim30$ My for an age of $\sim100$ My. However, the
Pleiades primordial cloud could have captured stars from a neighboring,
older region nearby (e.g., Bhatt 1989). Indeed, the age spread for the
Pleiades is comparable to that in whole OB associations or star
complexes, and the Pleiades, like most clusters, probably formed in such
a region.

\section{Implications}
\label{sect:impl}

The formation of stars in only one or two crossing times implies that
cloud lifetimes are short and the observed turbulent energy does not
have to be resupplied. Turbulent dissipation times are this short anyway
(Stone, Ostriker, \& Gammie 1998; MacLow, et al. 1998), so the
implication is that {\it all clouds proceed directly to star formation on a
dissipation time and never require rejuvenation or self-sustaining
feedback}. Fine-tuning of cloud stability from feedback should be very
difficult anyway, since protostellar wind speeds are much larger than
cloud escape speeds, and the wind energy should just escape through
fractal holes and tunnels (see also Henning 1989). 

Short timescales also imply that protostars do not have time to orbit
inside their cloud cores. For example, Palla \& 
Stahler (1999) noted that the stars in Orion could not have 
moved very far from their birthsites. 
Each star essentially stays where its initial
clump first became unstable, and it does not move around to interact
with other gas or distant protostars (although it may interact with one
or two near neighbors).   
Maps of self-gravitating protostellar clumps by
Motte, Andr\'e, \& Neri (1998) and Testi \& Sargent (1998) illustrate
this point: the protostars in the Ophiuchus and Serpens cores have such
small individual angular filling factors that each one would have to
orbit many times (the inverse of this filling factor multiplied by the
relative gravitational cross section) in the cloud core to interact with
each other. This result would seem to rule out models of the IMF based
on clump or protostar interactions, such as those by Price \&
Podsiadlowski (1995), Allen \& Bastien (1995), Murray \& Lin (1996),
Bonnell et al. (1997), Bonnell, Bate, \& Zinnecker (1998), and others.
Instead, IMF models based on the availability of gas to make stars in an
overall fractal network seem preferred (Elmegreen 1997a, 1999).

As a result of this birthsite freeze-out, the {\it youngest} star
positions should appear fractal, or hierarchical, like the gas in which
they form (see reviews in Elmegreen \& Efremov 1999; Elmegreen et al.
2000).
Larson (1995) and Simon (1997) discussed power-law two-point
correlation functions for star fields, but this is not necessarily
the same as a fractal distribution, and the fields they studied were
probably too old (Bate, Clarke \& McCaughrean 1998; Nakajima et al. 1998).
Gomez et al. (1993) discussed hierarchical structure in Taurus, which
is more to the point. A recent study by Vavrek, Bal\'azs, \& Epchtein
(1999) finds multifractal structure in young star positions.

Short cloud lifetimes have implications for chemistry too. Most chemical
reactions should be occurring at the high density of a
turbulent-compressed clump, which may be around $10^5$ cm$^{-3}$ (e.g.,
Falgarone, Phillips \& Walker 1991; Lada, Evans \& Falgarone 1997),
rather than the low average density that is observed in studies with
poor angular resolution. The formation of some chemical species at
elevated temperatures in turbulent shocks has already been noted
(Falgarone et al. 1995; Joulain et al. 1998).

The rapid rate of star formation suggested here for individual clouds does
not imply there should be a rapid rate of star formation on a galactic
scale, as suggested by Zuckerman \& Evans (1974) and Zuckerman \& Palmer
(1974).  Even if star formation proceeds on a dynamical timescale, the
actual time depends on the size of the cloud that contains it.  This is
true even on a galactic scale, where the star formation rate is generally
a fixed fraction of the density (i.e., $\epsilon\rho$ for ``efficiency''
$\epsilon$ and density $\rho$) divided by the local orbit time (Elmegreen
1997b; Kennicutt 1998). This galactic rate probably involves the same
physical principles that apply to individual complexes, associations and
clusters, all of which form stars at a rate equal to the {\it local}
$\epsilon\rho$ divided by the {\it local} dynamical time. There is no
catastrophe in the galactic star formation rate if all regions evolve on
a dynamical time, because the dynamical time is very long on a galactic
scale.

The essential point is that star formation does not occur in every
location where the gas is dense, it occurs primarily in self-gravitating
cores that comprise only a small fraction of the total cloud mass. Even
if most of a cloud is in the form of extremely dense clumps (Falgarone
1989), because of turbulence compression for example, most of these
clumps are generally stable and unable to form stars (e.g., Bertoldi \&
McKee 1992; Falgarone, Puget, \& P\'erault 1992). Some gas is probably
in a lower density interclump medium too, which is also unable to form
stars. Thus a lot of gas contributes to the total CO emission in our
Galaxy, but it does not contribute to star formation.

\newpage
\begin{figure}
\vspace{6.in}
\includegraphics{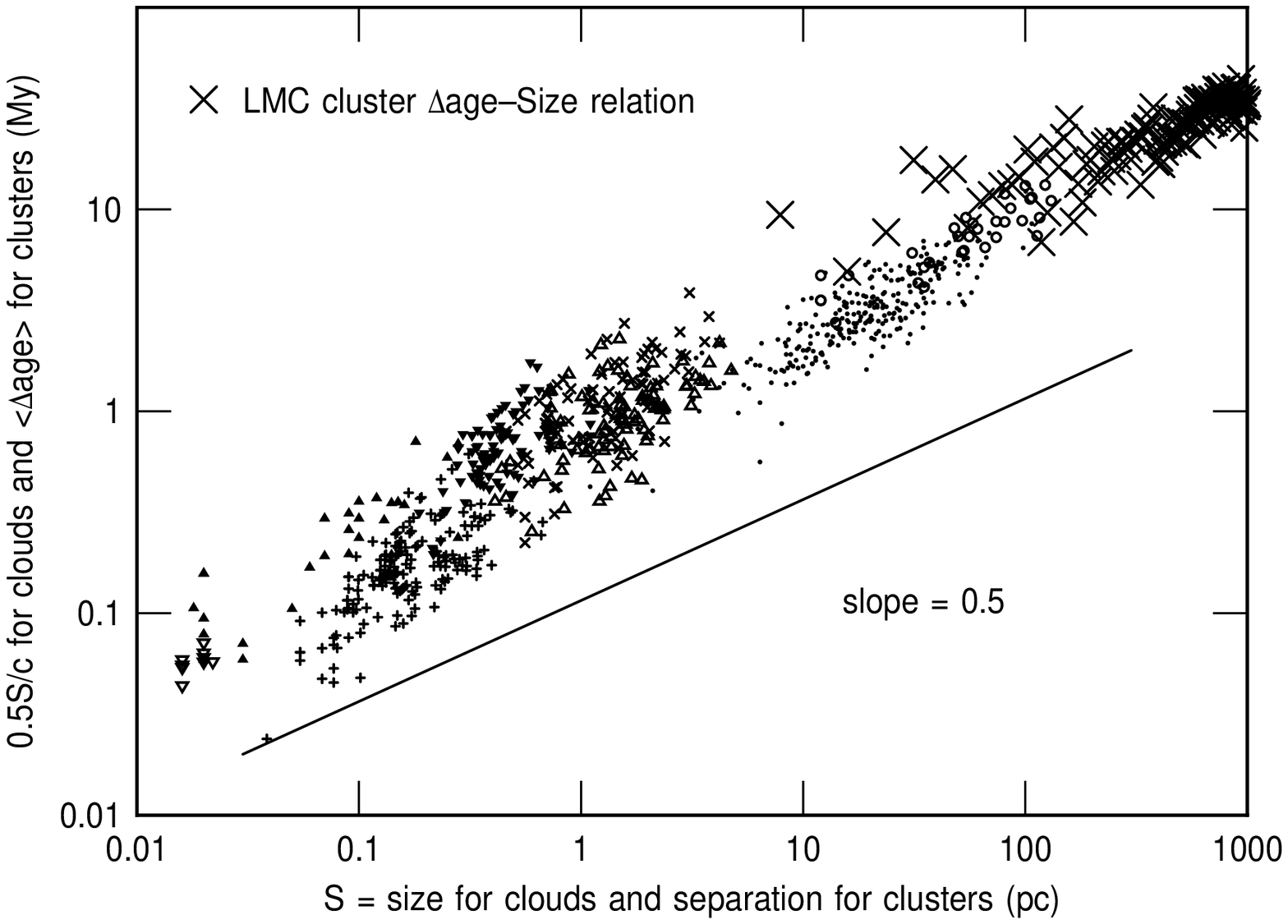}
\caption{Crossing time versus size for molecular clouds and their
cores (small symbols) and the average age difference versus 
separation for clusters in the LMC (large crosses; from 
Efremov \& Elmegreen 1998; see review in Elmegreen 
et al. 2000).  The different symbols represent different
cloud surveys: {\it dots:} Solomon et al. 1987; {\it open circles:}
Dame et al. 1986; {\it filled triangles:} Falgarone et al. 1992;
{\it open triangles:} Maddalena-Thaddeus cloud in 
Williams, et al. (1994); {\it upside
down open triangles:} Lemme et al. (1995);
{\it upside down filled triangles:} Loren (1989);
{\it pluses:} Stutzki \& Gusten (1990);
{\it crosses:} Rosette in Williams et al. (1994.)}

\end{figure}
\end{document}